\documentclass[journal]{IEEEtran}

\usepackage{amsmath,amssymb,amsfonts}
\usepackage{graphicx}
\usepackage{booktabs}
\usepackage{multirow}
\usepackage{array}
\usepackage{tabularx}
\usepackage{siunitx}
\usepackage{cite}
\usepackage{url}
\usepackage{hyperref}
\usepackage{algorithm}
\usepackage{algpseudocode}
\usepackage{tikz}
\usepackage{enumitem}
\usepackage{xcolor}
\usepackage{subcaption}
\usepackage{color}
\usepackage[normalem]{ulem}
\usepackage{soul}
\usepackage{eso-pic}
\usetikzlibrary{positioning,fit,calc,arrows.meta}
\usetikzlibrary{arrows.meta,positioning,shapes.geometric,fit,calc}
\usepackage{tabularx}  
\usepackage{booktabs} 

\hypersetup{
    colorlinks=true,
    linkcolor=black,
    citecolor=black,
    urlcolor=blue
}

\newcommand{\PUE}{\mathrm{PUE}}

\newcommand{\PIT}{P_{\mathrm{IT}}}

\newcommand{\AcceptedNotice}{%
  \AddToShipoutPictureBG*{%
    \AtPageUpperLeft{%
      \put(\LenToUnit{0.5\paperwidth},\LenToUnit{-1.0cm}){%
        \makebox[0pt][c]{\parbox{0.85\paperwidth}{\centering\footnotesize
          Accepted for publication at the 2027 IEEE PES Grid Edge Technologies
          Conference \& Exposition, Salt Lake City, UT, USA, April 19--22, 2027.}}}}%
    \AtPageLowerLeft{%
      \put(\LenToUnit{0.5\paperwidth},\LenToUnit{0.8cm}){%
        \makebox[0pt][c]{\parbox{0.85\paperwidth}{\centering\scriptsize
          \copyright~2027 IEEE. Personal use of this material is permitted.
          Permission from IEEE must be obtained for all other uses, in any current
          or future media, including reprinting/republishing this material for
          advertising or promotional purposes, creating new collective works, for
          resale or redistribution to servers or lists, or reuse of any copyrighted
          component of this work in other works.}}}}%
  }%
}
\title{Scalable Reference Electrical Architectures for Modern Data Centers}
\author{Kiran Kumar Challa, \IEEEmembership{Senior Member, IEEE}, Yan (Joann)  Zhou, and Kumarsinh Jhala, \IEEEmembership{Senior Member, IEEE}  

\thanks{K. K. Challa, Yan (Joann)  Zhou, and Kumarsinh Jhala are with Argonne National Laboratory, Lemont, IL, USA
(e-mail: kchalla@anl.gov, yzhou@anl.gov, kjhala@anl.gov).}
 \thanks{The submitted manuscript has been created by UChicago Argonne, LLC,
Operator of Argonne National Laboratory (“Argonne”) for Department of Energy under Contract No. DE-AC02-06CH11357. The work is funded by Office of Technology Commercialization and Analysis and Strategy Office of DOE.}%

}

\begin{document}
\AcceptedNotice
\maketitle

\begin{abstract}
The rapid growth of cloud computing, artificial intelligence (AI), and high-performance computing is driving the deployment of data centers with electrical demands ranging from tens of megawatts to multi-gigawatt campuses, creating significant challenges for utility interconnection and power system planning. This paper presents scalable, vendor-neutral reference electrical architectures for colocation, hyperscale, and AI data centers. The proposed framework integrates modular Power PODs, hierarchical HV/MV/LV distribution, standby generation, uninterruptible power supplies (UPS), battery energy storage systems (BESS), and scalable utility interfaces into a unified planning methodology. Three representative Power POD configurations are developed to support systematic capacity expansion while maintaining consistent electrical topology and equipment sizing. The proposed architectures facilitate utility interconnection studies, equipment demand forecasting, transmission planning, and grid infrastructure assessment, providing an open and reproducible benchmark for evaluating the impacts of large-scale data center deployment on future power systems.
\end{abstract}


\section{Introduction}
\label{sec:introduction}

\IEEEPARstart{M}{odern} data centers have evolved from enterprise facilities with electrical capacities of only a few megawatts into hyperscale and AI campuses requiring hundreds of megawatts, with multiple announced developments approaching or exceeding the gigawatt scale
\cite{IEA2025}. This unprecedented growth has made
data centers one of the fastest-growing classes of large electrical loads,
introducing significant challenges for electric utilities, transmission owners,
system operators, regulators, and equipment manufacturers
\cite{Mitra2023}. This has also intensified concerns
regarding the availability of critical grid-interconnection equipment \cite{DOEGridSupply2024}. Long
manufacturing lead times for large power transformers, gas-insulated
switchgear, medium-voltage switchgear, standby generators, and other electrical
equipment have become significant constraints on both utility expansion and
data-center deployment
\cite{LBNL2024}. Consequently, planning
studies increasingly require not only accurate forecasts of electrical demand
but also reproducible methodologies for estimating the associated electrical
infrastructure required to support projected data-center growth \cite{NERC2025Characteristics}.

Electrical infrastructure architectures (reference designs) published by industrial organizations have
become widely adopted during the conceptual and detailed design of modern data
centers. Guidance documents from Schneider Electric \cite{SchneiderRD101}, Vertiv \cite{VertivPowerTrain}, NVIDIA \cite{NvidiaGB200Reference}, the
Uptime Institute \cite{UptimeTierTopology}, and the ISO/IEC~22237 series \cite{ISO22237} provide valuable recommendations
regarding power distribution topologies, redundancy philosophies, modular
deployment strategies, and facility reliability. These
references have significantly improved engineering practice by standardizing
many aspects of facility design. However, these documents are primarily intended for engineering implementation
and frequently emphasize vendor-specific solutions or reference installations.
They generally do not provide an open, technology-neutral methodology that can
be directly incorporated into transmission planning studies, utility
interconnection analyses, infrastructure forecasting, or reproducible academic
research.

From the power-system perspective, the rapid growth of hyperscale and AI data centers has emerged as a major planning challenge, driving extensive research on transmission expansion, resource adequacy, voltage stability, congestion management, dynamic performance, and large-load interconnection studies \cite{Mitra2023,LBNL2024}. Utilities, independent system operators, and regional transmission organizations have recognized that future planning requires more detailed representations of data-center electrical infrastructure to accurately assess grid impacts, equipment requirements, and utility interconnection strategies. However, most existing power-system studies represent data centers as aggregated constant-power loads or simplified demand profiles for steady-state and planning analyses, without explicitly modeling the underlying electrical infrastructure \cite{JimenezRuiz2025}. Such simplified representations are often inadequate for capturing the dynamic behavior and grid interactions of modern data centers, motivating the development of physics-based electrical models for planning and stability studies \cite{Chalamala2025,ESM2026}. Furthermore, available reference electrical architectures are largely proprietary or vendor-specific, limiting transparency, reproducibility, and their applicability to academic research and utility planning \cite{SchneiderRD101,VertivPowerTrain}. Consequently, there remains no standardized, scalable, and vendor-neutral electrical architecture that bridges facility-level design with grid level studies.

To address these challenges, this paper presents a scalable, vendor-neutral reference electrical architecture for modern data centers that bridges facility-level electrical design with utility, transmission, and distribution planning. In this paper, ‘vendor-neutral’ denotes a modular, technology-agnostic planning methodology that uses functional categories and standardized equipment sizing and topology, ensuring direct applicability across different manufacturers and open compatibility with industry standards. The proposed framework integrates utility substations, HV/MV/LV distribution, modular power POD, standby generation, UPS, BESS, and IT infrastructure into a unified architecture suitable for colocation, hyperscale, and AI data centers. Based on established industry standards and reference designs \cite{ISO22237,TIA942,SchneiderRD101,UptimeTierTopology}, the framework provides a reproducible benchmark for power-system planning, utility interconnection studies, and infrastructure expansion. The main contributions of this work are summarized as follows:

\begin{itemize}

\item A scalable and vendor-neutral reference electrical architecture for colocation, hyperscale, and AI data centers that supports both facility design and grid planning.

\item A modular power POD methodology with standardized HV, MV, and LV electrical topologies enables consistent capacity expansion and utility interconnection studies.

\item A reproducible planning framework for evaluating electrical infrastructure requirements, equipment demand, and future transmission and distribution planning associated with large-scale data center deployment.

\end{itemize}

The remainder of this paper describes the proposed reference architectures, the design methodology, representative planning applications, and the concluding remarks.

\section{Scalable Reference Electrical Architectures}
\label{sec:architecture}

The framework represents the principal
functional elements found in modern data-center electrical systems: the utility
service interface; high-voltage (HV) substations; HV/medium-voltage (MV)
transformers; MV switchgear and feeders; modular power PODs; standby generators;
automatic transfer switches (ATSs); uninterruptible power-supply (UPS) systems;
battery energy-storage systems (BESSs); low-voltage (LV) switchboards; power
distribution units (PDUs). These elements
are consistent with the functional building blocks used in commercial reference
designs and with reliability-oriented data-center topology standards
\cite{SchneiderReferenceDesigns,UptimeTierTopology,ISO22237}.

\begin{figure}[!h]
\centering
\resizebox{\columnwidth}{!}{%
\begin{tikzpicture}[
    node distance=0.42cm,
    every node/.style={font=\footnotesize},
    block/.style={rectangle, rounded corners, draw=black, thick,
                  align=center, minimum width=2.45cm, minimum height=0.75cm,
                  fill=blue!10},
    hvblock/.style={rectangle, rounded corners, draw=black, thick,
                    align=center, minimum width=2.45cm, minimum height=0.75cm,
                    fill=purple!12},
    PODblock/.style={rectangle, rounded corners, draw=black, thick,
                     align=center, minimum width=2.45cm, minimum height=0.75cm,
                     fill=green!12},
    loadblock/.style={rectangle, rounded corners, draw=black, thick,
                      align=center, minimum width=2.2cm, minimum height=0.70cm,
                      fill=orange!15},
    line/.style={-Latex, thick}
]

\node[block] (grid) {Utility Transmission Grid\\(Representative 69--345~kV)};
\node[hvblock, below=of grid] (sub) {HV Substation(s)\\Protection, Switching, and Metering};
\node[hvblock, below=of sub] (hvmv) {HV/MV Transformer Bank(s)};
\node[hvblock, below=of hvmv] (mv) {Campus MV Switchgear\\and MV Distribution};

\node[PODblock, below left=0.55cm and 1.35cm of mv] (POD1)
{Power POD 1\\MV/LV Transformer\\ATS--UPS--BESS};
\node[PODblock, below=0.55cm of mv] (POD2)
{Power POD 2\\MV/LV Transformer\\ATS--UPS--BESS};
\node[PODblock, below right=0.55cm and 1.35cm of mv] (PODn)
{Power POD $N$\\MV/LV Transformer\\ATS--UPS--BESS};

\node[loadblock, below=of POD1] (load1) {IT, Cooling,\\and Auxiliary Loads};
\node[loadblock, below=of POD2] (load2) {IT, Cooling,\\and Auxiliary Loads};
\node[loadblock, below=of PODn] (loadn) {IT, Cooling,\\and Auxiliary Loads};

\node[loadblock, right=0.7cm of sub] (highmw)
{High-Capacity Expansion\\Additional HV Bays, Transformers,\\Switchgear, and Reactive Support};

\draw[line] (grid) -- (sub);
\draw[line] (sub) -- (hvmv);
\draw[line] (hvmv) -- (mv);
\draw[line] (mv) -- (POD1);
\draw[line] (mv) -- (POD2);
\draw[line] (mv) -- (PODn);
\draw[line] (POD1) -- (load1);
\draw[line] (POD2) -- (load2);
\draw[line] (PODn) -- (loadn);
\draw[line, dashed] (sub.east) -- (highmw.west);

\end{tikzpicture}}
\caption{Overall reference electrical architecture for scalable modern data
centers. }
\label{fig:overall_architecture}
\end{figure}
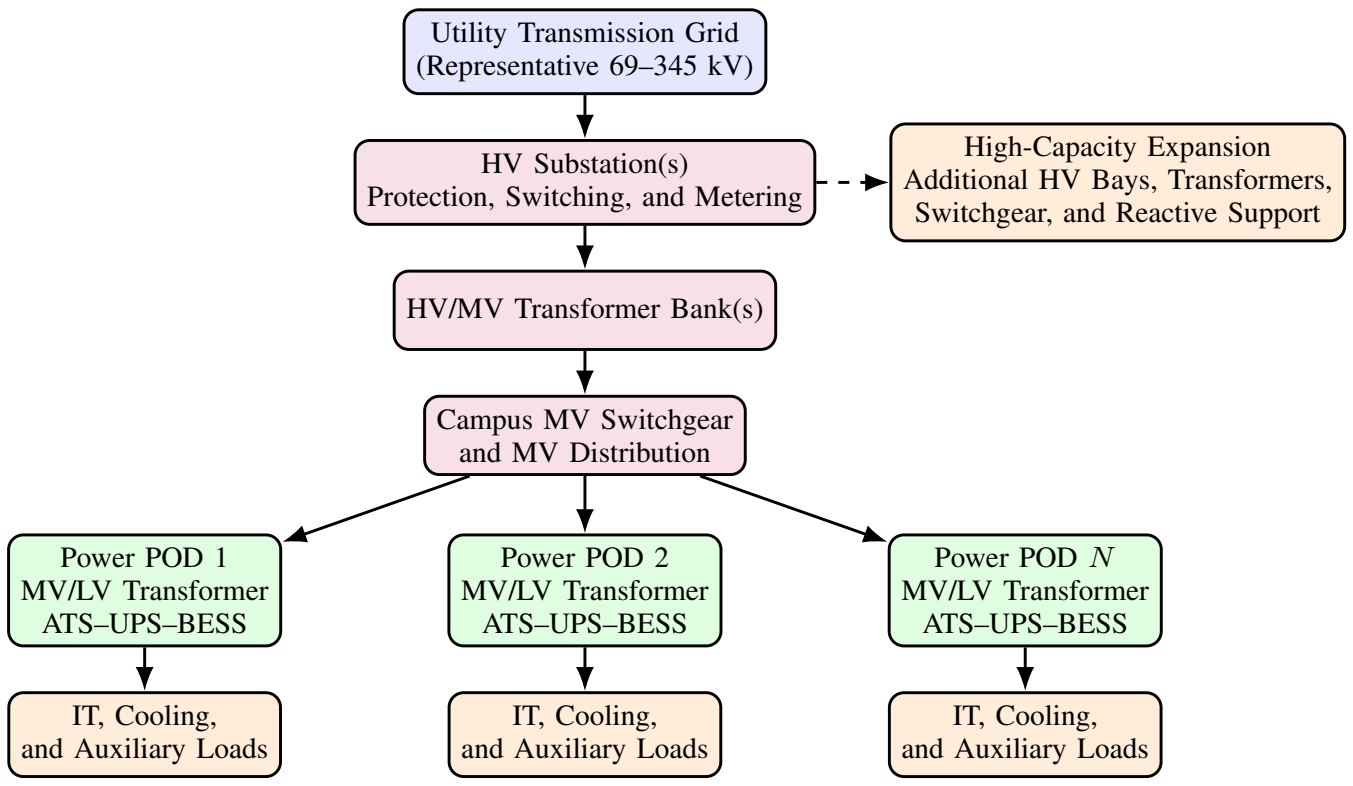

The architecture follows a hierarchical and modular design, as shown in Fig.~\ref{fig:overall_architecture}. The architecture separates electrical infrastructure into HV, MV, and LV layers
to preserve a consistent planning representation over a wide range of facility
sizes. It also reflects the modular scaling
principles adopted in contemporary cloud and accelerated-computing reference
architectures \cite{SchneiderReferenceDesigns,NvidiaGB200Reference}. The proposed end-to-end
architecture is discussed in the following subsections.

\subsection{Utility Interface and High-Capacity Expansion}
The facility is supplied from the utility transmission or
subtransmission system through one or more dedicated HV substations. The utility interface includes revenue metering, protection, disconnecting devices, busbars, and the HV/MV transformer banks required to establish the campus distribution voltage. Data center facilities with \(<100\ MW\) can be integrated through a single utility substation or a
limited number of transformer banks, whereas large campuses exceeding \(100\ MW\) require multiple utility level
substations, transmission-line extensions, or dedicated network upgrades. Utility interface can be expanded in terms of a \(100\ MW\) modular blocks \cite{VertivPowerTrain}. Each block adds substation bays, transformer capacity, HV and MV switchgear, and
associated protection and control while retaining the same downstream POD
structure.

Along with block structure for \(>100\ MW\); the interconnection study must also evaluate contingency
performance, voltage regulation, reactive-power capability, short-circuit duty,
harmonic performance, and simultaneous loss or restoration of large load blocks \cite{NERC2026RiskMitigation,NERC2025Characteristics}. Utilities do install compensating devices such as static synchronous compensators (STATCOMs), synchronous condensers, switched
capacitors, or other reactive-power resources when the interconnection studies identify inadequate dynamic voltage support. Data centers also install facility level battery energy storage to smooth out it's load variations and improve the reliability at facility level.

\subsection{HV, MV, and LV Distribution Layers}

At the MV layer, switchgear, bus sections, feeders, and protection devices
distribute power to individual PODs. Dedicated POD feeders allow selective
isolation and phased construction, while sectionalized buses and alternate
sources can provide maintenance flexibility. The specific MV voltage and feeder
topology are selected based on developer practice, facility capacity, voltage drop, and equipment availability. 
The MV system distributes power through dedicated feeders to independent power
PODs. Dual-corded IT equipment can be supplied from independent
paths where the required reliability objective warrants it. However, redundancy
must be assessed for the entire end-to-end topology rather than inferred from a
single duplicated component \cite{UptimeTierTopology,ISO22237}.

\subsection{Modular Power-POD Architecture}

\begin{figure}[!h]
    \centering
    \includegraphics[width=0.6\linewidth]{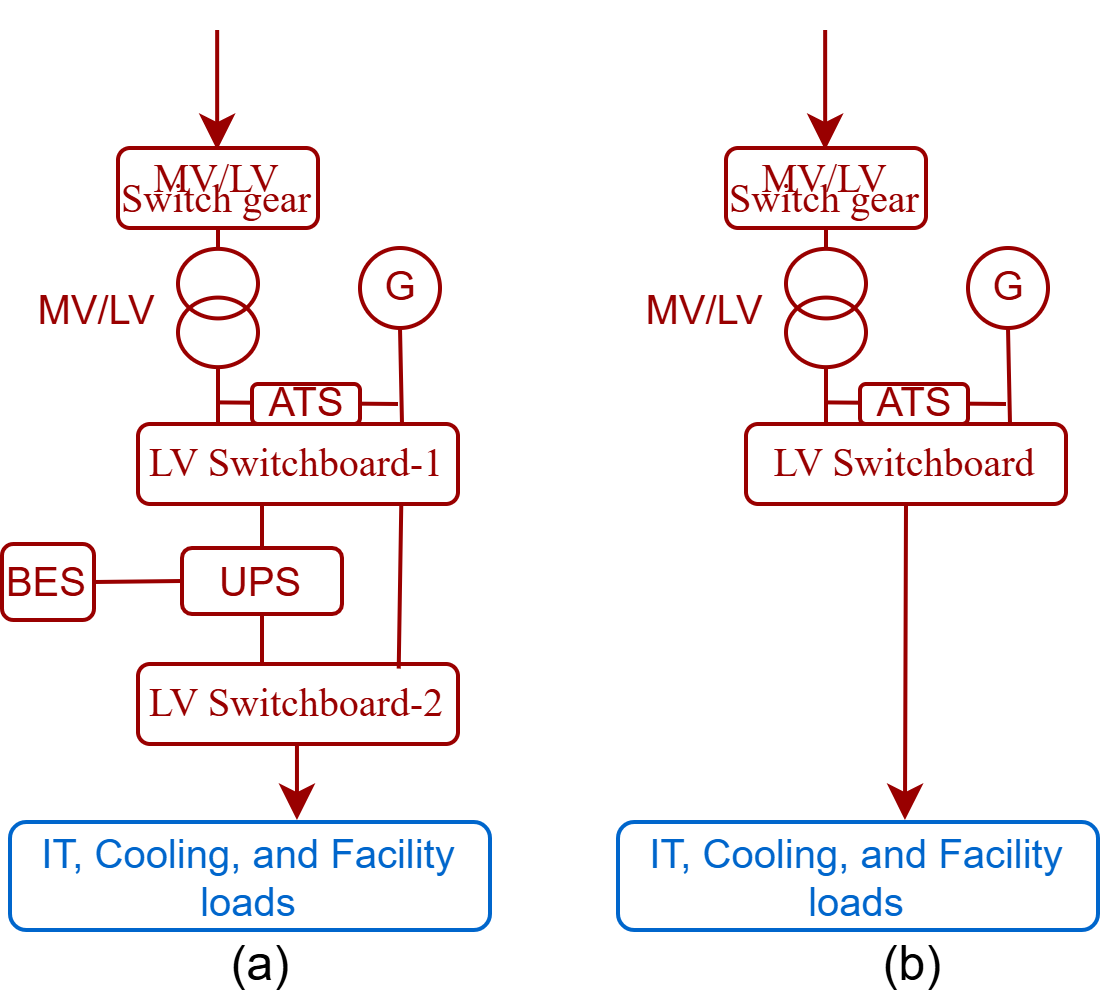}
    \caption{Simplified reference power-POD electrical architecture.}
    \label{fig:POD_architecture}
\end{figure}

The power POD serves as the fundamental modular and repeatable building block of modern data center electrical architectures. Major infrastructure vendors, including Schneider Electric and Vertiv, adopt POD-based electrical designs that enable standardized deployment, modular expansion, and scalable capacity growth, while hyperscale operators similarly organize facilities into repeatable capacity blocks for incremental expansion \cite{SchneiderAI2026,Vertiv360AI2025}. It supplies a defined portion of IT capacity together with the
corresponding cooling and auxiliary demand. A representative POD shown in Fig. \ref{fig:POD_architecture} includes an
MV/LV transformer, LV switchboard, ATS, standby generator, UPS system, battery
bank, PDU or busway system, and downstream rack distribution. 

During normal operation, the preferred utility source supplies the POD through
the MV/LV transformer and LV switchboard. The UPS conditions power to critical
IT loads and provides ride-through using its battery system during short-duration
voltage disturbances or source interruptions. If the disturbance persists, the
standby generator starts and the transfer system connects the supported bus to
the emergency source. The topology therefore distinguishes instantaneous
ride-through from longer-duration standby generation, consistent with common
high-availability data-center practice \cite{UptimeTierTopology,NFPA110_2022,IEC62040_3}.

The reference architecture separates IT, cooling, and auxiliary loads, enabling explicit PUE modeling and flexibility in cooling technologies. This is particularly important for AI/HPC facilities, where high-density computing tightly integrates power and thermal infrastructure \cite{NvidiaGB200Reference}. Cooling, LV distribution, and auxiliary requirements are therefore tailored to colocation, hyperscale, and AI data centers. Detailed component design and assumptions are discussed in the following section.

\section{Design architecture and Engineering Assumptions}
\label{sec:design_methodology}
The proposed reference architecture is intended for planning-level studies
rather than detailed facility engineering. The objective is to translate a
small set of facility characteristics into a consistent electrical
representation of a data center that can be used for utility interconnection
analysis, transmission and distribution planning, equipment-demand estimation,
and comparative assessment of future data-center deployments.

The architecture is generated from four primary inputs summarized in Table. \ref{tab:planning_inputs}: the facility IT
capacity, power POD IT capacity, power usage effectiveness (PUE), and
data-center class. The IT capacity defines the overall scale of the facility,
while the selected POD capacity determines the modular building blocks used to
construct the electrical architecture. The PUE relates the IT demand to the
total facility demand, and the data-center class determines the representative
POD capacities and PUE values used by the planning methodology. These inputs
are sufficient to generate the complete electrical representation, including
POD-level equipment, campus electrical infrastructure, and utility
interconnection requirements.

\begin{table}[!h]
\caption{Primary Planning Inputs}
\label{tab:planning_inputs}
\centering
\footnotesize
\begin{tabular}{p{0.30\columnwidth}p{0.16\columnwidth}p{0.42\columnwidth}}
\toprule
\textbf{Input} & \textbf{Symbol} & \textbf{Description} \\
\midrule
IT Capacity & $\PIT$ &
Total rated IT load of the facility \\

Power POD Capacity & $P_{\mathrm{pod}}$ &
IT capacity assigned to one power POD \\

Power Usage Effectiveness & $\PUE$ &
Ratio of total facility demand to IT demand \\

Data Center Class & -- &
Colocation, Hyperscale, or AI/HPC (determines default POD capacity and PUE) \\
\bottomrule
\end{tabular}
\end{table}

\subsection{Utility Interconnection and HV/MV Infrastructure Planning}
The campus HV/MV infrastructure is sized based on the facility apparent-power capacity, $S_{\mathrm{facility}}$, and provides the interface between the utility supply and power PODs \cite{ISO22237,UptimeTierTopology}. The methodology is applicable to commonly used MV distribution voltages (e.g., \SIlist{12.47;13.2;13.8;22;34.5}{\kilo\volt}) by selecting equipment with appropriate voltage and power ratings \cite{IEEE3001_2}. As facility capacity increases, additional HV/MV transformers, substations, and switchgear are introduced for scalability and redundancy according to Table~\ref{tab:hv_selection}.

\begin{table}[!h]
\caption{Reference HV/MV Infrastructure Planning}
\label{tab:hv_selection}
\centering
\footnotesize
\begin{tabular}{ccccc}
\toprule
Facility &
HV &
HV/MV &
Transformer &
MV \\
Capacity (MVA) &
Subs. &
XFMRs &
Rating (MVA) &
SWGR \\
\midrule
$\leq15$      & 0 & 0 & --   & 0 \\
$15$--$30$    & 2 & 2 & $S_{\mathrm{facility}}$   & 2 \\
$30$--$80$    & 3 & 3 & $S_{\mathrm{facility}}/2$ & 3 \\
$>80$         & 4 & 4 & $S_{\mathrm{facility}}/3$ & 4 \\
\bottomrule
\end{tabular}
\end{table}

For large data centers, dedicated transmission-level interconnection equipment may be required \cite{NERC2025Characteristics,NERC2026RiskMitigation}. The proposed methodology represents this infrastructure in \SI{100}{\mega\watt} IT-capacity blocks, with
\[
N_{\mathrm{block}}=\left\lceil\frac{P_{\mathrm{IT}}}{100}\right\rceil,
\qquad
N_{\mathrm{utility}}=N_{\mathrm{block}}+1,
\]
where the additional unit provides a planning allowance for redundancy and expansion. These quantities represent dedicated substations, transformers, and associated switchyard equipment beyond the campus infrastructure in Table~\ref{tab:hv_selection}. For very large campuses, STATCOM sizing is treated separately as a planning-level estimate of dynamic reactive-power support at the grid interface. Two discrete STATCOM ratings, \SI{106}{\mega\volt\ampere} and \SI{64}{\mega\volt\ampere}, are considered. Defining

\begin{equation}
N_{500}^{S}=
\mathrm{INT}\left(\frac{S_{\mathrm{facility}}}{500}\right),
\qquad
R=S_{\mathrm{facility}}-500N_{500}^{S},
\end{equation}

where $R$ is the remaining facility capacity in MVA, the number of \SI{106}{\mega\volt\ampere} STATCOM units is

\begin{equation}
Q_{106}=
N_{500}^{S}+
\begin{cases}
1, & R>350,\\
0, & \text{otherwise},
\end{cases}
\end{equation}

and the number of \SI{64}{\mega\volt\ampere} STATCOM units is

\begin{equation}
Q_{64}=
\begin{cases}
1, & 250\leq R\leq350,\\
0, & \text{otherwise}.
\end{cases}
\end{equation}
with the final compensation requirement determined through utility interconnection and system studies.

\subsection{Power-POD-Based Equipment Estimation}

The proposed methodology adopts the power POD as the fundamental modular
building block of the data center electrical infrastructure. The required
number of PODs is determined from the target IT capacity as $N_{POD}=\frac{P_{IT}}{P_{POD}}$, where $P_{\mathrm{IT}}$ is the desired IT capacity and
$P_{\mathrm{pod}}$ is the selected POD IT capacity. Since PODs are deployed as
discrete modules, the installed IT capacity may exceed the requested capacity,
providing reserve capacity for future expansion and phased deployment.

Each POD has a predefined apparent-power rating determined by its IT capacity
and the selected power usage effectiveness (PUE). The reference POD ratings
used in this work are summarized in Table~\ref{tab:pod_mva_lookup}. The total
installed facility apparent-power capacity is therefore $S_{\mathrm{facility}}=N_{\mathrm{pod}}S_{\mathrm{pod}}$, where $S_{\mathrm{pod}}$ is the apparent-power rating of an individual POD.

\begin{table}[!h]
\caption{Reference POD Apparent-Power Ratings}
\label{tab:pod_mva_lookup}
\centering
\begin{tabular}{c|ccc}
\toprule
\multirow{2}{*}{POD IT Capacity} &
\multicolumn{3}{c}{POD Rating (MVA)}\\
\cline{2-4}
 & PUE 1.10 & PUE 1.20 & PUE 1.30\\
\midrule
1.5 MW & x & 2.5 & 2.5\\
2.5 MW & 3.5 & 4.0 & 4.0\\
5.0 MW & 7.0 & 7.5 & x\\
\bottomrule
\end{tabular}
\end{table}

The electrical equipment associated with each POD is standardized to enable
consistent estimation across different data center scales. The reference
quantities adopted in this work are listed in
Table~\ref{tab:sample_facilities}. Consequently, the quantity of any equipment
type is obtained as $N_x=q_xN_{\mathrm{pod}}$, where $q_x$ is the number of units assigned to each POD for equipment type
$x$. The base architecture consists of one MV/LV transformer, one standby
generator, one automatic transfer switch (ATS), one LV switchboard, four UPS
modules, four battery cabinets, and twelve PDUs per POD.

The proposed framework represents the base modular architecture and can be
extended to accommodate different redundancy philosophies (e.g., $N$, $N+1$,
or $2N$) by modifying the equipment quantities. Standby generators provide
backup power during sustained utility outages, while UPS systems and battery
storage supply the short-duration ride-through required until standby
generation becomes available. This modular representation provides a scalable
and vendor-independent basis for estimating electrical equipment requirements
across a wide range of modern data center architectures.

\subsection{Architecture Generation Procedure}
The proposed methodology generates the data center electrical architecture in a
hierarchical manner, beginning with the power POD, followed by the campus
electrical infrastructure, and finally the utility interconnection. The overall
procedure is summarized in Algorithm~\ref{alg:architecture_generation}.

\begin{algorithm}[!h]
\caption{Data Center Architecture Generation}
\label{alg:architecture_generation}
\begin{algorithmic}[1]
\Require $P_{\mathrm{IT}}$, $P_{\mathrm{pod}}$, PUE, data-center class
\Ensure Electrical architecture and equipment quantities
\State Determine the required number of power PODs
\State Select the POD apparent-power rating
\State Compute the installed facility capacity
\State Estimate POD-level equipment quantities
\State Determine campus HV/MV infrastructure
\If{$P_{\mathrm{IT}}>100$ MW}
    \State Add utility-interconnection equipment
\EndIf
\If{$P_{\mathrm{IT}}\geq300$ MW}
    \State Add STATCOM installations
\EndIf
\State Apply the selected redundancy philosophy
\State Generate the complete equipment summary
\end{algorithmic}
\end{algorithm}

The architecture is generated progressively across three hierarchical levels.
First, the target IT capacity is partitioned into standardized power PODs,
which serve as the fundamental modular building blocks of the facility. Each
POD contributes a predefined set of electrical equipment, including MV/LV
transformers, standby generators, ATSs, UPS modules, battery systems, and
power distribution units. The aggregate POD capacity is then used to determine
the required campus electrical infrastructure, including HV/MV substations,
transformer banks, and MV switchgear.

\section{Sample Calculation}
The proposed scalable architecture methodology is illustrated using \SI{100}{\mega\watt} and \SI{500}{\mega\watt} IT facilities with \SI{2.5}{\mega\watt} power PODs and six PODs per data hall. The required numbers of PODs and data halls are

\begin{equation}
N_{\mathrm{POD}}=
\left\lceil\frac{P_{\mathrm{IT}}}{P_{\mathrm{POD}}}\right\rceil,
\qquad
N_{\mathrm{hall}}=
\left\lceil\frac{N_{\mathrm{POD}}}{N_{\mathrm{POD/hall}}}\right\rceil.
\end{equation}

\begin{figure}[!h]
    \centering
    \includegraphics[width=1.0\linewidth]{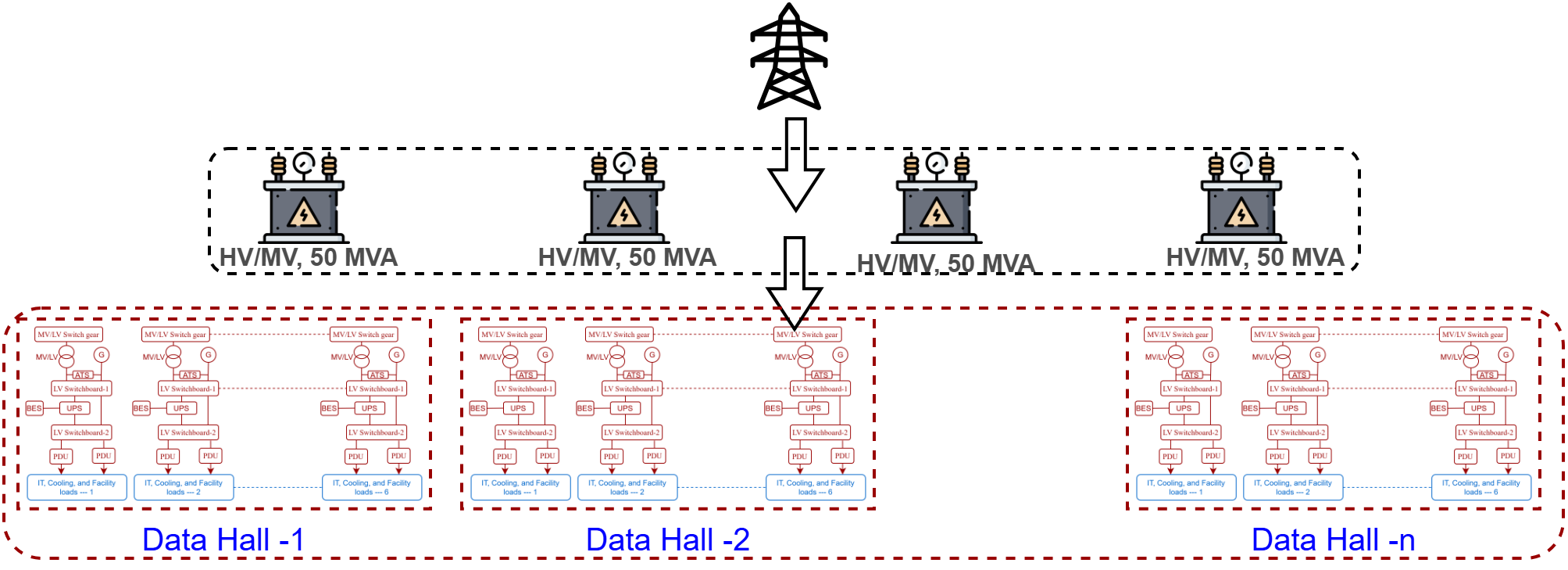}
    \caption{Architecture of data center with 100 MW IT capacity}
    \label{100MW_architecture}
\end{figure}

For the \SI{100}{\mega\watt} facility, $N_{\mathrm{POD}}=40$ and $N_{\mathrm{hall}}=7$. Based on the reference equipment quantities per POD, the facility requires 40 MV/LV transformers, standby generators, ATSs, and LV switchboards; 160 UPS modules and battery cabinets/strings; and 480 PDUs. At the facility level, a \SI{100}{\mega\watt} IT block is assumed to require approximately \SI{150}{\mega\volt\ampere} of transformation capacity. This is provided by three \SI{50}{\mega\volt\ampere} HV/MV transformer sets plus one redundant set, resulting in four transformer sets.

For the \SI{500}{\mega\watt} facility,
\begin{equation}
N_{\mathrm{POD}}=\frac{500}{2.5}=200,
\qquad
N_{\mathrm{hall}}=
\left\lceil\frac{200}{6}\right\rceil=34.
\end{equation}

Thus, the facility contains 200 power PODs distributed across 34 data halls, requiring 200 MV/LV transformers, standby generators, ATSs, and LV switchboards; 800 UPS modules and battery cabinets/strings; and 2400 PDUs. The campus is organized as five \SI{100}{\mega\watt} modules with an aggregate transformation requirement of approximately \SI{750}{\mega\volt\ampere}. At the campus level, three \SI{250}{\mega\volt\ampere} HV/HV transformers provide the required capacity, with a fourth unit providing N+1 redundancy.

For a facility apparent-power capacity of \SI{500}{\mega\volt\ampere} STATCOM requirements are,
\begin{equation}
N_{500}^{S}=\mathrm{INT}\left(\frac{500}{500}\right)=1,
\qquad R=0,
\end{equation}

giving $Q_{106}=1$ and $Q_{64}=0$. Therefore, one \SI{106}{\mega\volt\ampere} STATCOM is included in the \SI{500}{\mega\volt\ampere} reference design. This rating is intended for planning-level equipment estimation; detailed reactive-power compensation should be determined through utility interconnection, power-flow, contingency, and dynamic voltage-stability studies.

\begin{table}[!t]
\caption{Sample Equipment Requirements}
\label{tab:sample_facilities}
\centering
\footnotesize
\setlength{\tabcolsep}{2.0pt}
\begin{tabular}{lccc}
\toprule
\textbf{Equipment} &
\textbf{Per POD} &
\textbf{100 MW} &
\textbf{500 MW} \\
\midrule
Power PODs               & -- & 40  & 200  \\
Data halls               & -- & 7   & 34   \\
MV/LV transformers       & 1  & 40  & 200  \\
Standby generators       & 1  & 40  & 200  \\
ATS                      & 1  & 40  & 200  \\
LV switchboards          & 1  & 40  & 200  \\
UPS modules              & 4  & 160 & 800  \\
Battery cabinets/strings & 4  & 160 & 800  \\
PDUs                     & 12 & 480 & 2400 \\
\midrule
50-MVA HV/MV XFMRs       & -- & 4   & 20 \\
250-MVA HV/HV XFMRs      & -- & --  & 4 \\
106-MVA STATCOM          & -- & --  & 1 \\
\bottomrule
\end{tabular}
\end{table}

Table~\ref{tab:sample_facilities} summarizes the resulting equipment requirements from the power-POD level to the campus grid interface.The examples demonstrate the hierarchical scaling of the reference architecture. POD-level equipment scales with installed IT capacity, while the upstream infrastructure transitions from \SI{50}{\mega\volt\ampere} HV/MV transformer sets for a \SI{100}{\mega\watt} module to an aggregated \SI{250}{\mega\volt\ampere} HV/HV interface for the \SI{500}{\mega\watt} campus. STATCOMs are added independently based on facility apparent-power capacity to represent planning-level grid-interface reactive-power support.

Overall, the framework scales with facility size by adding infrastructure at predefined capacity thresholds. Since all equipment requirements originate from the modular power-POD architecture, the same methodology can be consistently applied across data-center types and capacities, providing a scalable and reproducible approach for equipment-demand estimation and preliminary infrastructure planning.

\section{Conclusions and Future Work}
\label{sec:conclusion}

This paper presented a scalable, vendor-neutral reference electrical architecture for modern colocation, hyperscale, and AI data centers. The proposed framework combines a hierarchical electrical distribution architecture with modular power PODs and capacity-based planning rules to estimate facility electrical infrastructure and utility interconnection requirements. By explicitly representing substations, transformers, switchgear, standby generation, battery energy storage, and reactive power support, the methodology provides a reproducible benchmark for equipment-demand estimation, utility interconnection studies, and long-term transmission and distribution planning. The modular and hierarchical architecture enables consistent representation of facilities ranging from tens of megawatts to multi-gigawatt campuses while remaining adaptable to different redundancy philosophies, utility standards, and regional design practices. 

Future work will extend the benchmark by incorporating time-varying workload models, detailed dynamic models of on-site energy resources and microgrid controllers, and integrated transmission–distribution representations for evaluating large-load interconnections and grid impacts. 



\bibliographystyle{IEEEtran}
\bibliography{references}

@techreport{NERC2026RiskMitigation,
  author       = {{North American Electric Reliability Corporation (NERC)}},
  title        = {{Reliability Guideline: Risk Mitigation for Emerging Large Loads}},
  institution  = {NERC},
  address      = {Atlanta, GA, USA},
  month        = may,
  year         = {2026},
  type         = {Reliability Guideline},
}

@techreport{NERC2025Characteristics,
  author       = {{North American Electric Reliability Corporation (NERC)}},
  title        = {{Characteristics and Risks of Emerging Large Loads}},
  institution  = {NERC},
  address      = {Atlanta, GA, USA},
  month        = jul,
  year         = {2025},
  type         = {White Paper},
}

@techreport{DOEGridSupply2024,
  author       = {{U.S. Department of Energy}},
  title        = {{2021--2024 Four-Year Review of Supply Chains for the Energy Sector Industrial Base}},
  institution  = {U.S. Department of Energy},
  address      = {Washington, DC, USA},
  year         = {2024},
  month        = dec,
  type         = {Technical Report},
}

@manual{UptimeTierTopology,
  author       = {{Uptime Institute}},
  title        = {{Tier Standard: Topology}},
  organization = {Uptime Institute},
  edition      = {2018},
  address      = {New York, NY, USA},
  year         = {2018},
  url          = {https://airs.uptimeinstitute.com/resources/asset/tier-standard-topology},
  note         = {Accessed: Jul. 23, 2026}
}

@techreport{ESM2026,
  author       = {Babu Chalamala and others},
  title        = {Data Centers for Future Power Systems},
  institution  = {IEEE Energy Sustainability Magazine},
  year         = {2026},
  month        = feb,
  volume       = {3},
  number       = {1},
  pages        = {56--67}
}

@techreport{Chalamala2025,
  author       = {Babu Chalamala and Veronika Rabl and others},
  title        = {Data Center Growth and Grid Readiness},
  institution  = {IEEE Power \& Energy Society},
  type         = {Technical Report TR131},
  year         = {2025},
  month        = may,
  doi          = {10.17023/W4WY-S557}
}

@article{JimenezRuiz2025,
  author  = {Alberto Jimenez-Ruiz and Federico Milano},
  title   = {Data Center Model for Transient Stability Analysis of Power Systems},
  journal = {IEEE Transactions on Power Systems},
  year    = {2025},
  note    = {Early Access}
}

@techreport{SchneiderAI2026,
  author       = {{Schneider Electric}},
  title        = {{How 6 AI Attributes Change Data Center Design}},
  institution  = {Schneider Electric},
  type         = {White Paper},
  number       = {WP110 Version 3.1},
  year         = {2026},
  month        = may,
  url          = {https://www.se.com/ww/en/download/document/SPD_WP110_EN/},
  note         = {Accessed: Jul. 23, 2026}
}

@techreport{Vertiv360AI2025,
  author       = {{Vertiv}},
  title        = {Vertiv 360AI High Density Reference Design \#027A: 2.5 MW PODs at up to 142 kW per Rack},
  institution  = {Vertiv},
  type         = {Reference Design},
  year         = {2025},
  month        = mar,
  url          = {https://www.vertiv.com/4936a8/globalassets/campaigns/ai-hub/overhaul/reference-design/vertiv-reference-design-027a_datasheet_apac-en_2025_sl-80148.pdf},
  note         = {Accessed: Jul. 23, 2026}
}

@standard{NFPA110_2022,
  author       = {{National Fire Protection Association}},
  title        = {{NFPA 110: Standard for Emergency and Standby Power Systems}},
  organization = {National Fire Protection Association},
  number       = {NFPA 110},
  edition      = {2022},
  address      = {Quincy, MA, USA},
  year         = {2022},
}

@misc{VertivPowerTrain,
  author       = {{Vertiv}},
  title        = {{The Data Center Power Train: Managing Energy from Grid to Chip}},
  year         = {2025},
  month        = mar,
  organization = {Vertiv},
  type         = {E-book},
  address      = {Westerville, OH, USA},
  url          = {https://www.vertiv.com/492f32/globalassets/shared/vertiv-datacenterpowertrain-eb-en-na-sl-80203-web-small.pdf},
  note         = {Accessed: Jul. 24, 2026}
}

@standard{IEC62040_3,
  author       = {{International Electrotechnical Commission}},
  title        = {{IEC 62040-3:2021 Uninterruptible Power Systems (UPS) -- Part 3: Method of Specifying the Performance and Test Requirements}},
  organization = {International Electrotechnical Commission},
  number       = {IEC 62040-3:2021},
  edition      = {3.0},
  address      = {Geneva, Switzerland},
  year         = {2021},
  month        = apr,
}

@misc{NvidiaGB200Reference,
  author       = {{NVIDIA Corporation}},
  title        = {{NVIDIA Contributes NVIDIA GB200 NVL72 Designs to Open Compute Project}},
  year         = {2024},
  month        = oct,
  howpublished = {\url{https://developer.nvidia.com/blog/nvidia-contributes-nvidia-gb200-nvl72-designs-to-open-compute-project/}},
  note         = {NVIDIA Technical Blog, Accessed: Jul. 24, 2026}
}

@standard{IEEE3001_2,
  author       = {IEEE Std 3001.2-2017},
  title        = {{Recommended Practice for Evaluating the Electrical Service Requirements of Industrial and Commercial Power Systems}},
  organization = {Institute of Electrical and Electronics Engineers},
  address      = {Piscataway, NJ, USA},
  year         = {2017},
  month        = nov,
  doi          = {10.1109/IEEESTD.2017.8048872},
}

@standard{ISO22237,
  author       = {},
  title        = {{ISO/IEC 22237-1:2021, Information Technology---Data Centre Facilities and Infrastructures---Part 1: General Concepts}},
  organization = {International Organization for Standardization and International Electrotechnical Commission},
  address      = {Geneva, Switzerland},
  year         = {2021},
}

@standard{TIA942,
  title={Telecommunications Infrastructure Standard for Data Centers},
  organization={Telecommunications Industry Association},
  number={ANSI/TIA-942},
  year={2024}
}

@techreport{LBNL2024,
  author = {{Arman Shehabi et al.}},
  title = {{2024 United States Data Center Energy Usage Report}},
  institution = {Lawrence Berkeley National Laboratory},
  address = {Berkeley, CA, USA},
  number = {LBNL-2001637},
  year = {2024},
  month = dec,
  doi = {10.71468/P1WC7Q},
  url = {https://doi.org/10.71468/P1WC7Q},
  note = {Prepared for the U.S. Department of Energy, Office of Energy Efficiency and Renewable Energy (EERE), Office of Industrial Efficiency and Decarbonization (IEDO), under Contract No. DE-AC02-05CH11231. Accessed: Jul. 24, 2026}
}

@techreport{IEA2025,
  author={{International Energy Agency}},
  title={{Energy and AI}},
  institution={IEA},
  year={2025}
}

@misc{SchneiderRD101,
  author       = {{Schneider Electric}},
  title        = {{Reference Design 101: 56 MW Tier III Data Center Reference Design}},
  year         = {2024},
  organization = {Schneider Electric},
  url          = {https://www.se.com/us/en/work/solutions/data-centers-and-networks/reference-designs/},
  urldate      = {2026-07-24}
}

@misc{SchneiderReferenceDesigns,
  author={{Schneider Electric}},
  title={EcoStruxure Data Center Reference Designs},
  year={2025}
}

@article{Mitra2023,
  author       = {Parag Mitra},
  title        = {Transmission Planning and Large Data Centers},
  journal      = {IEEE Electrification Magazine},
  volume       = {11},
  number       = {3},
  pages        = {90--92},
  month        = sep,
  year         = {2023},
  publisher    = {Institute of Electrical and Electronics Engineers (IEEE)},
  doi          = {10.1109/MELE.2023.3291509},
  issn         = {2325-5897},
}

\end{document}